\documentclass[aps,prd,showpacs,showkeys,twocolumn,10pt]{revtex4}

\usepackage{amssymb}
\usepackage{amsmath}
\pdfoutput=1
\usepackage[breaklinks]{hyperref}   % hipernuorodoms: \href{url}{text}

\usepackage[T1]{fontenc}
\usepackage{multirow}

\begin{document}

\title{Radiative M1 transitions of heavy baryons in the bag model}


\author{A.~Bernotas$^{1}$ and V.~\v{S}imonis$^{2}$
\vspace*{1.2\baselineskip}}

\affiliation{$^1$ Vilnius
University Faculty of Physics, Saul\.{e}tekio 9, LT-10222 Vilnius, Lithuania
\vspace*{1.2\baselineskip} \\
$^2$ Vilnius University Institute of Theoretical Physics and
Astronomy, A. Go\v{s}tauto 12, LT-01108,\ Vilnius, Lithuania\\}

\date{\today}

\begin{abstract}
We study the M1 transitions of ground state heavy baryons within a framework
of the modified bag model. Calculations of transition moments and
corresponding M1 decay widths are performed. For the $\mathrm{spin\,}\frac{1%
}{2}$ baryons containing three differently flavoured quarks the hyperfine
mixing effects are taken into account. Results are compared with
estimates obtained using various other approaches.

\end{abstract}

\maketitle

\section{\label{sec_int}Introduction}

A study of electromagnetic properties of baryons plays an important role in
the elementary particle physics. Electromagnetic observables serve as a
source of information on the structure of hadrons. In our recent paper \cite
{BS13} we have used the modified bag model to calculate magnetic
moments of $J=\frac{1}{2}$ and $J=\frac{3}{2}$ charmed and bottom baryons.
For the baryons made of three differently flavoured quarks the
colour-hyperfine mixing was taken into account. In the present paper we
continue our exploration of heavy baryon electromagnetic structure through
the calculation of their M1 transition moments. We seek to give some
estimates for the radiative decay rates of heavy baryons as well. We expect it
to be a useful step towards the comprehensive description of heavy baryon
properties. Radiative decays of doubly heavy baryons have been studied (including
hyperfine mixing effects) in nonrelativistic potential model \cite{AHN10}
and in relativistic three-quark model \cite{BFG10}. We are going to compare
our corresponding predictions with the results obtained in these papers.

The format of our paper is as follows. In Sec.~\ref{sec_bag} we give a brief
overview of the model and present basic expressions necessary for our
investigation. The results of our calculations for transition moments and
decay widths are presented in Sec.~\ref{sec_cal} and Sec.~\ref{sec_rad},
respectively. The last section is a short summary.

\section{\label{sec_bag}Bag model and transition magnetic moments}

The model we use to calculate the transition magnetic
moments is exactly the same as has been used in our previous work \cite{BS13}%
. For completeness we remind here the main features of this model (for
details we refer to \cite{BS04}),
%with accent on the differences between our variant of the model and 
emphasizing its differences from the original MIT bag one \cite{DJJK75}.

The energy of the bag associated with a particular hadron is given by 
%2.01
\begin{equation}
E=\frac{4\pi }{3}BR^{3}+\frac{Z_{0}}{R}+\sum\limits_{i}\varepsilon
_{i}+\Delta E\,,  \label{eq2.01}
\end{equation}
where $R$ denotes the bag radius, and the four terms on the right-hand side
of this expression are: the bag volume energy, the Casimir energy, the sum
of single-particle eigenenergies, and the quark-quark interaction energy due to
one-gluon-exchange. The bag radius $R_{\mathrm{H}}$ of each hadron is
obtained by minimizing (\ref{eq2.01}) with respect to $R$.

We use the effective strong coupling constant and effective (running) quark
mass. They are defined as 
%2.02-2.03
\begin{eqnarray}
\alpha _{\mathrm{c}}(R) & = & \frac{2\pi }{9\ln (A+R_{0}/R)}\;, 
\label{eq2.02} \\
\overline{m}_{f}(R) & = & \widetilde{m}_{f}+\alpha _{\mathrm{c}}(R)\cdot \delta
_{f}\,.  
\label{eq2.03}
\end{eqnarray}

The bag energy corrected for the center-of-mass motion (c.m.m.) is
identified with the mass of hadron. It is related to the uncorrected one by
%2.04
\begin{equation}
M^{2}=E^{2}-P^{2},  
\label{eq2.04}
\end{equation}
where 
%2.05
\begin{equation}
P^{2}=\gamma \sum\limits_{i}p_{i}^{2}\,  
\label{eq2.05}
\end{equation}
is the effective momentum square, and $p_{i}=\sqrt{\varepsilon _{i}-m_{i}}$
represent momenta of individual quarks.

The c.m.m. corrected magnetic moments are given by the relation 
%2.06
\begin{equation}
\mu =\frac{E}{M}\ \mu ^{0}.  
\label{eq2.06}
\end{equation}

The model parameters are: the bag constant $B$, the Casimir energy parameter 
$Z_{0}$, the parameter governing the c.m.m. prescription $\gamma $, two
parameters from the definition of the running coupling constant ($A$ and $%
R_{0}$), and six parameters necessary to define the mass functions (Eq.~\ref
{eq2.03}) for the strange, charmed, and bottom quarks ($\widetilde{m}_{f}$, $%
\delta _{f}$). Light ($u$ and $d$) quarks are assumed to be massless. All
parameters are the same as in our previous work~\cite{BS13} ($B=7.468\times
% \times, kaip IX lenteleje 
10^{-4}~\mathrm{GeV}^{4}$, $Z_{0}=0.22$, $\gamma =2.153$, $A=0.6514$, $%
R_{0}=4.528~\mathrm{GeV}^{-1}$, $\widetilde{m}_{s}=0.262~\mathrm{GeV}$, $%
\delta _{s}=0.083$~$\mathrm{GeV}$, $\widetilde{m}_{c}=1.458~\mathrm{GeV}$, $%
\delta _{c}=0.089~\mathrm{GeV}$, $\widetilde{m}_{b}=4.721~\mathrm{GeV}$, and 
$\delta _{b}=0.079~\mathrm{GeV}$).

The magnetic moment of a quark confined in the bag of radius $R_{\mathrm{B}}$
is given by 
%2.07-2.08
\begin{eqnarray}
\mu _{i} & = & q_{i}\ \bar{\mu}_{i}\ ,  
\label{eq2.07} \\
\bar{\mu}_{i} & = & \frac{4\varepsilon _{i}R_{\mathrm{B}}+2m_{i}R_{\mathrm{B}}-3}{%
2(\varepsilon _{i}R_{\mathrm{B}}-1)\varepsilon _{i}R_{\mathrm{B}}+m_{i}R_{%
\mathrm{B}}}\ \frac{R_{\mathrm{B}}}{6}\ ,  
\label{eq2.08}
\end{eqnarray}
where $q_{i}$ is the electric charge of the quark, and $\bar{\mu}_{i}$
denotes reduced (charge-independent) magnetic moment (see \cite{DJJK75}).

The wave functions of ground state baryons can be constructed by coupling
the spins of the two first quarks to an intermediate spin $S$ and then
adding the third one to obtain the total spin $J$': 
%2.09
\begin{subequations}
\begin{align}
\left| B\right\rangle & = \left| (q_{1}q_{2})^{S=0}q_{3},J=\frac{1}{2}%
\right\rangle \,, 
\label{eq2.09a} \\
\left| B^{\prime }\right\rangle & = \left| (q_{1}q_{2})^{S=1}q_{3},\,J=\frac{1}{%
2}\right\rangle \,,
\label{eq2.09b} \\
\left| B^{*}\right\rangle & = \left| (q_{1}q_{2})^{S=1}q_{3},J=\frac{3}{2}%
\right\rangle \,.
\label{eq2.09c}
\end{align}
\end{subequations}

The valence quark contribution to the baryon magnetic moments is \cite
{FLNC81} 
%2.10
\begin{subequations}
\begin{align}
\mu (B) & =\mu _{3}\,, 
\label{eq2.10a} \\
\mu (B^{\prime }) & =\frac{1}{3}(2\mu _{1}+2\mu _{2}-\mu _{3})\,, 
\label{eq2.10b} \\
\mu (B^{*}) & =\mu _{1}+\mu _{2}+\mu _{3}\,,
\label{eq2.10c}
\end{align}
\end{subequations}
where $\mu _{i}$ denote the magnetic moments of first, second, and third
quarks, respectively. For the transition magnetic moments we have 
%2.11
\begin{subequations}
\begin{align}
\mu (B^{\prime }\leftrightarrow B) & =\frac{1}{\sqrt{3}}(\mu _{2}-\mu _{1})\,,
\label{eq2.11a} \\
\mu (B^{*}\leftrightarrow B) & =\sqrt{\frac{2}{3}}(\mu _{1}-\mu _{2})\,,
\label{eq2.11b} \\
\mu (B^{*}\leftrightarrow B^{\prime }) & =\frac{\sqrt{2}}{3}(\mu _{1}+\mu
_{2}-2\mu _{3})\,.
\label{eq2.11c}
\end{align}
\end{subequations}

The signs of transition moments depend on the adopted phase convention. Ours
coincide with the one from Ref.~\cite{FLNC81}.

Using Eqs.~(\ref{eq2.10a})--(\ref{eq2.10c}) and (\ref{eq2.11a})--(\ref{eq2.11c}) the $\mathrm{spin\,}\frac{3}{2}%
\leftrightarrow \frac{1}{2}$ transition moments can be expressed in terms of
others: 
%2.12-2.13
\begin{eqnarray}
\mu (B^{*}\leftrightarrow B^{\prime }) & = & \frac{\sqrt{2}}{3}\left[ 3\mu
(B^{\prime })-\mu (B^{*})\right] \,, 
\label{eq2.12} \\
\mu (B^{*}\leftrightarrow B) & = & -\sqrt{2}\mu (B^{\prime }\leftrightarrow B)\,.
\label{eq2.13}
\end{eqnarray}

The baryons containing three quarks of different flavours need a special
treatment. In this case the intermediate spin $S$ is no longer a good
quantum number. The colour-hyperfine interaction mixes the states with
different intermediate spins, so that physical states are linear
combinations of initial ones: 
%2.14
\begin{subequations}
\begin{align}
\left| B_{\text{phys}}\right\rangle & =C_{1}\left| B\right\rangle +C_{2}\left|
B^{\prime }\right\rangle \,, 
\label{eq2.14a} \\ 
\left| B_{\text{phys}}^{\prime }\right\rangle & =-C_{2}\left| B\right\rangle
+C_{1}\left| B^{\prime }\right\rangle \,. 
\label{eq2.14b}
\end{align}
\end{subequations}

The physical (mixed) magnetic and transition moments are 
%2.15
\begin{subequations}
\begin{align}
\mu (B_{\text{phys}}) & =C_{1}^{2}\,\mu (B)+C_{2}^{2}\,\mu (B^{\prime
})+2C_{1}C_{2}\,\mu (B^{\prime }\leftrightarrow B)\,, 
\label{eq2.15a}\\ 
\mu (B_{\text{phys}}^{\prime }) & =C_{1}^{2}\,\mu (B^{\prime })+C_{2}^{2}\,\mu
(B)-2C_{1}C_{2}\,\mu (B^{\prime }\leftrightarrow B)\,,
\label{eq2.15b}
\end{align}
\end{subequations}
%2.16
\begin{eqnarray}
\mu (B_{\text{phys}}^{\prime } \leftrightarrow
B_{\text{phys}}) & = & (C_{1}^{2}\,-C_{2}^{2}\,)\,\mu (B^{\prime }\leftrightarrow B)
\nonumber \\
&&{}+C_{1}C_{2}[\,\mu (B^{\prime })-\mu (B)]\,,  
\label{eq2.16}
\end{eqnarray}
%2.17
\begin{subequations}
\begin{align}
\mu (B_{\text{phys}}^{*}\leftrightarrow B_{\text{phys}}) & = C_{1}\mu (B^{*}\leftrightarrow
B)+C_{2}\mu (B^{*}\leftrightarrow B^{\prime })\,, 
\label{eq2.17a}\\ 
\mu (B_{\text{phys}}^{*}\leftrightarrow B_{\text{phys}}^{\prime }) & = C_{1}\mu
(B^{*}\leftrightarrow B^{\prime })-C_{2}(B^{*}\leftrightarrow B)\,.
\label{eq2.17b}
\end{align}
\end{subequations}

The mixing of states is the reason that Eqs.~(\ref{eq2.12}) and (\ref{eq2.13}%
), as they stand, are practically useless, because the states $\left|
B\right\rangle $ and $\left| B^{\prime }\right\rangle $ in general are not
the physical states. These relations are valid only in the cases when the
state mixing is absent or, at least, very small. Throughout, we work in the
limit of exact isospin symmetry and thus neglect the small $\Sigma _{Q}-\Lambda
_{Q}$ mixing. Other states unaffected by the hyperfine mixing are the
remaining members of $\Sigma _{Q}$ isomultiplet and the states corresponding
to baryons $\Omega _{Q}$, $\Xi _{QQ}$, $\Omega _{QQ}$, where $Q$ denotes
heavy flavours ($c$ and $b$). Such are also the ground states of triply
heavy baryons $\Omega _{bcc}^{+}$ and $\Omega _{bbc}^{0}$. For all of them
Eq.~(\ref{eq2.12}) holds. Note that they all are of $B^{\prime }$ type
(economy in primes leads to some mess-up in notations). As concerns Eq.~(\ref
{eq2.13}), for the charmed baryon $\Sigma _{c}^{+}$ and bottom baryon $%
\Sigma _{b}^{0}$ (in the limit of exact isospin symmetry) we have 
%2.18
\begin{equation}
\mu (\Sigma _{Q}^{*}\leftrightarrow \Lambda _{Q})=-\sqrt{2}\mu (\Sigma
_{Q}\leftrightarrow \Lambda _{Q})\,,  \label{eq2.18}
\end{equation}
where $\Sigma _{Q}=(\Sigma _{c}^{+},$ $\Sigma _{b}^{0})$, and $\Lambda
_{Q}=(\Lambda _{c}^{+},\Lambda _{b}^{0})$. In this case $\Lambda _{Q}$ is of 
$B$ type, and $\Sigma _{Q}$ is of $B^{\prime }$ type.

Isospin symmetry also leads to some additional relations:
%2.19
\begin{subequations}
\begin{align}
\mu (\Sigma _{c}^{*\,+}\leftrightarrow \Sigma _{c}^{+}) & = \frac{1}{2}[\mu
(\Sigma _{c}^{*\,0}\leftrightarrow \Sigma _{c}^{0})+\mu (\Sigma
_{c}^{*\,++}\leftrightarrow \Sigma _{c}^{++})]\,, 
\label{eq2.19a}\\ 
\mu (\Sigma _{c}^{*\,0}\leftrightarrow \Sigma _{c}^{0}) & = -\frac{2\sqrt{2}}{3}%
\mu (\Sigma _{c}^{*+})\,, 
\label{eq2.19b}\\ 
\mu (\Sigma _{c}^{*\,++}\leftrightarrow \Sigma _{c}^{++}) & = -\frac{2\sqrt{2}}{3%
}\mu (\Sigma _{c}^{*0})\,, 
\label{eq2.19c}\\ 
\mu (\Xi _{cc}^{*\,+}\leftrightarrow \Xi _{cc}^{+}) & = \frac{\sqrt{2}}{3}\mu
(\Xi _{cc}^{*++})\,,
\label{eq2.19d}
\end{align}
\end{subequations}
%2.20
\begin{subequations}
\begin{align}
\mu (\Sigma _{b}^{*\,0}\leftrightarrow \Sigma _{b}^{0}) & = \frac{1}{2}[\mu
(\Sigma _{b}^{*\,-}\leftrightarrow \Sigma _{b}^{-})+\mu (\Sigma
_{b}^{*\,+}\leftrightarrow \Sigma _{b}^{+})]\,, 
\label{eq2.20a}\\ 
\mu (\Sigma _{b}^{*-}\leftrightarrow \Sigma _{b}^{-}) & = -\frac{2\sqrt{2}}{3}%
\mu (\Sigma _{b}^{*0})\,, 
\label{eq2.20b}\\ 
\mu (\Sigma _{b}^{*\,+}\leftrightarrow \Sigma _{b}^{+}) & = -\frac{2\sqrt{2}}{3}%
\mu (\Sigma _{b}^{*-})\,, 
\label{eq2.20c}\\ 
\mu (\Xi _{bb}^{*\,-}\leftrightarrow \Xi _{bb}^{-}) & = \frac{\sqrt{2}}{3}\mu
(\Xi _{bb}^{*0})\,.
\label{eq2.20d}
\end{align}
\end{subequations}

Above we considered relations involving only the states with the same quark
content. The naive quark model offers a somewhat richer collection of various
relations including the states with different quark content, as, e.\,g., 
%2.21-2.22
\begin{eqnarray}
\mu (\Omega _{cc}^{*\,+}\leftrightarrow \Omega _{cc}^{+}) & \approx & -\mu
(\Omega _{c}^{*\,0}\leftrightarrow \Omega _{c}^{0})\,,  \label{eq2.21} \\
\mu (\Omega _{bcc}^{*\,+}\leftrightarrow \Omega _{bcc}^{+}) & \approx & -\mu
(\Omega _{bbc}^{*\,0}\leftrightarrow \Omega _{bbc}^{0})\,,  \label{eq2.22}
\end{eqnarray}
etc. However, in the bag model the magnetic moments of light ($u,d$),
strange, and (to some extent) charmed quarks are sensitive to the
environment in which they reside. Therefore the accuracy of such relations
is very low. Equations (\ref{eq2.21}) and (\ref{eq2.22}) represent some kind of
exception -- the accuracy of the former (in our variant of the bag model) is $\approx 6\%$ and of the latter $%
\approx 3\%$. For all other relations of this type the accuracy is worse
than $10\%$.

The relations presented above are a nice manifestation of underlying symmetry,
however, in the absence of experimental data, at the time being we cannot
check their validity in practice.

\section{\label{sec_cal}Calculation of $\mathrm{spin\,}\frac{3}{2}%
\leftrightarrow \frac{1}{2}$ transition moments}

In this section we calculate the $\mathrm{spin\,}\frac{3}{2}\leftrightarrow 
\frac{1}{2}$ transition magnetic moments of all ground state heavy baryons.
Calculations are performed in the framework of modified bag model described
in the previous section. Only the valence quark contribution is taken into
account. The magnetic moments of quarks in the bag model depend on the bag
radius of the hadron under consideration. Therefore in the case of
transition moments a prescription which radius to use is necessary. One
may try to pick the smaller of the two with an intention to take into
account the overlap of bags. However, for heavy baryons both radii are
similar. We present the values of transition moments obtained using the
radii of lighter baryons ($B$). The opposite choice may cause a shift of
calculated values by less than $0.5\%$. Of course, such difference is
irrelevant. Also we need the prescription how to use the c.m.m. correction
(Eq.~\ref{eq2.06}) because the ratio $E/M$ for the baryons under transition
may differ. We have checked that for charmed and bottom baryons the
difference between $E_{B^{*}}/M_{B^{*}}$ and $E_{B}/M_{B}$ is sufficiently
small, and we can choose any of the two. Our prescription was to use the ratio $%
E_{B}/M_{B}$ corresponding to the lighter baryon again. The results of our
calculations are listed in Tables~\ref{t3.1} and \ref{t3.2}. We also compare
our predictions with some other theoretical estimates. These are:

\begin{itemize}
\item  Nonrelativistic quark model with screening and effective quark mass 
\cite{DV09} (SCR).

\item  Chiral constituent quark model \cite{SDCG10} ($\chi$CQM).

\item  Light cone QCD sum rules \cite{AAO09} (LCSR). The definition of
transition moment $G_{M}$ used in that paper differs from ours. The relation
between our results and theirs is $\mu =\sqrt{(3M_{B^{*}})/(2M_{B})}\left(
M_{P}/M_{B}\right) G_{M}$, where $M_{P}$ is the mass of the proton. The
factor $M_{P}/M_{B}$ is used to convert natural magneton into nuclear magneton.

\item  Simple nonrelativistic quark model (NRQM). Since we have found
few papers to compare our predictions with, we also give estimates obtained
in nonrelativistic quark model (with the state mixing accounted for the
baryons containing three differently flavoured quarks). We treat these
results as a kind of reference point. Input values for quark magnetic
moments (in nuclear magnetons $\mu _{N}$) and state mixing angles (in
radians) were taken
from Ref.~\cite{FLNC81}: $\mu _{u}=-2\mu _{d}$, $\mu _{d}=-0.93~\mu _{N}$, $%
\mu _{s}=-0.61~\mu _{N}$, $\mu _{c}=0.39~\mu _{N}$, $\mu _{b}=-0.06~\mu _{N}$%
, $\theta _{usc}=\theta _{dsc}=0.066$, $\theta _{usb}=\theta _{dsb}=0.017$, $%
\theta _{ucb}=\theta _{dcb}=0.13$, $\theta _{scb}=0.12$.
\end{itemize}

%t3.1
\begin{table*}[htb]
\caption{ Spin $\frac{3}{2}-\frac{1}{2}$ transition moments (in nuclear
magnetons) of charmed baryons.\label{t3.1}}
\begin{center} 
\begin{tabular}{lccccc}
\hline\hline
Transition & Our & NRQM & SCR \cite{DV09}$^{**}$ & $\chi$CQM \cite{SDCG10} & LCSR 
\cite{AAO09}$^{**}$ \\ \hline
$\Sigma _{c}^{*0}\leftrightarrow \Sigma _{c}^{0}$ & \multicolumn{1}{r}{$%
-1.030$} & \multicolumn{1}{r}{$-1.24$} & \multicolumn{1}{r}{$1.07$} & 
\multicolumn{1}{r}{$1.48\phantom{0}$} & \multicolumn{1}{r}{$0.24\pm 0.05$} \\%
[2pt] 
$\Sigma _{c}^{*+}\leftrightarrow \Sigma _{c}^{+}$ & \multicolumn{1}{r}{$%
-0.062$} & \multicolumn{1}{r}{$0.07$} & \multicolumn{1}{r}{$0.08$} & 
\multicolumn{1}{r}{$-0.003$} & \multicolumn{1}{r}{$0.57\pm 0.09$} \\[2pt] 
$\Sigma _{c}^{*+}\leftrightarrow \Lambda _{c}^{+}$ & \multicolumn{1}{r}{$%
1.700$} & \multicolumn{1}{r}{$2.28$} & \multicolumn{1}{r}{$2.15$} & 
\multicolumn{1}{r}{$2.40\phantom{0}$} & \multicolumn{1}{r}{$2.00\pm 0.53$} \\%
[2pt] 
$\Sigma _{c}^{*++}\leftrightarrow \Sigma _{c}^{++}$ & \multicolumn{1}{r}{$%
0.905$} & \multicolumn{1}{r}{$1.39$} & \multicolumn{1}{r}{$1.23$} & 
\multicolumn{1}{r}{$-1.37\phantom{0}$} & \multicolumn{1}{r}{$1.33\pm 0.38$}
\\[2pt] 
$\Xi _{c}^{*0}\leftrightarrow \Xi _{c}^{0}$ & \multicolumn{1}{r}{$-0.224$} & 
\multicolumn{1}{r}{$-0.33$} & \multicolumn{1}{r}{$0.18$} & 
\multicolumn{1}{r}{$-0.50\phantom{0}$} & \multicolumn{1}{r}{$0.22\pm 0.07$}
\\[2pt] 
$\Xi _{c}^{*0}\leftrightarrow \Xi _{c}^{\prime \,0}$ & \multicolumn{1}{r}{$%
-0.915$} & \multicolumn{1}{r}{$-1.07$} & \multicolumn{1}{r}{$0.99$} & 
\multicolumn{1}{r}{$1.24\phantom{0}$} & --- \\[2pt] 
$\Xi _{c}^{*+}\leftrightarrow \Xi _{c}^{+}$ & \multicolumn{1}{r}{$1.497$} & 
\multicolumn{1}{r}{$2.03$} & \multicolumn{1}{r}{$1.94$} & \multicolumn{1}{r}{%
$2.08\phantom{0}$} & \multicolumn{1}{r}{$1.93\pm 0.72$} \\[2pt] 
$\Xi _{c}^{*+}\leftrightarrow \Xi _{c}^{\prime \,+}$ & \multicolumn{1}{r}{$%
-0.089$} & \multicolumn{1}{r}{$0.09$} & \multicolumn{1}{r}{$0.17$} & 
\multicolumn{1}{r}{$-0.23\phantom{0}$} & --- \\[2pt] 
$\Omega _{c}^{*0}\leftrightarrow \Omega _{c}^{0}$ & \multicolumn{1}{r}{$%
-0.839$} & \multicolumn{1}{r}{$-0.94$} & \multicolumn{1}{r}{$0.90$} & 
\multicolumn{1}{r}{$0.96\phantom{0}$} & --- \\[2pt] 
$\Xi _{cc}^{*+}\leftrightarrow \Xi _{cc}^{+}$ & \multicolumn{1}{r}{$0.945$}
& \multicolumn{1}{r}{$1.24$} & \multicolumn{1}{r}{$1.06$} & 
\multicolumn{1}{r}{$-1.41\phantom{0}$} & --- \\[2pt] 
$\Xi _{cc}^{*++}\leftrightarrow \Xi _{cc}^{++}$ & \multicolumn{1}{r}{$-0.787$%
} & \multicolumn{1}{r}{$-1.39$} & \multicolumn{1}{r}{$1.35$} & 
\multicolumn{1}{r}{$1.33\phantom{0}$} & --- \\[2pt] 
$\Omega _{cc}^{*+}\leftrightarrow \Omega _{cc}^{+}$ & \multicolumn{1}{r}{$%
0.789$} & \multicolumn{1}{r}{$0.94$} & \multicolumn{1}{r}{$0.88$} & 
\multicolumn{1}{r}{$-0.89\phantom{0}$} & --- \\ \hline\hline
\multicolumn{6}{l}{$^{\ast *}$~Only absolute values $\left| \mu \right| $
are presented.}
\end{tabular}
\end{center}
\end{table*}

%t3.2
\begin{table}[tbp] \centering%
\caption{Spin $\frac{3}{2}-\frac{1}{2}$ transition moments (in nuclear
magnetons) of bottom baryons.\label{t3.2}}
\begin{tabular}{lccc}
\hline\hline
Transition & Our & NRQM & LCSR \cite{AAO09}$^{**}$ \\ \hline
$\Sigma _{b}^{*-}\leftrightarrow \Sigma _{b}^{-}$ & \multicolumn{1}{r}{$%
-0.504$} & \multicolumn{1}{r}{$-0.82$} & \multicolumn{1}{r}{$0.42\pm 0.14$}
\\[2pt] 
$\Sigma _{b}^{*0}\leftrightarrow \Sigma _{b}^{0}$ & \multicolumn{1}{r}{$%
0.345 $} & \multicolumn{1}{r}{$0.49$} & \multicolumn{1}{r}{$0.20\pm 0.08$} \\%
[2pt] 
$\Sigma _{b}^{*0}\leftrightarrow \Lambda _{b}^{0}$ & \multicolumn{1}{r}{$%
1.488$} & \multicolumn{1}{r}{$2.28$} & \multicolumn{1}{r}{$1.52\pm 0.58$} \\%
[2pt] 
$\Sigma _{b}^{*+}\leftrightarrow \Sigma _{b}^{+}$ & \multicolumn{1}{r}{$%
1.193 $} & \multicolumn{1}{r}{$1.81$} & \multicolumn{1}{r}{$0.83\pm 0.28$} \\%
[2pt] 
$\Xi _{b}^{*-}\leftrightarrow \Xi _{b}^{-}$ & \multicolumn{1}{r}{$-0.139$} & 
\multicolumn{1}{r}{$-0.26$} & \multicolumn{1}{r}{$0.18\pm 0.06$} \\[2pt] 
$\Xi _{b}^{*-}\leftrightarrow \Xi _{b}^{\prime \,-}$ & \multicolumn{1}{r}{$%
-0.415$} & \multicolumn{1}{r}{$-0.66$} & --- \\[2pt] 
$\Xi _{b}^{*0}\leftrightarrow \Xi _{b}^{0}$ & \multicolumn{1}{r}{$1.321$} & 
\multicolumn{1}{r}{$2.03$} & \multicolumn{1}{r}{$1.71\pm 0.60$} \\[2pt] 
$\Xi _{b}^{*0}\leftrightarrow \Xi _{b}^{\prime \,0}$ & \multicolumn{1}{r}{$%
0.392$} & \multicolumn{1}{r}{$0.61$} & --- \\[2pt] 
$\Omega _{b}^{*-}\leftrightarrow \Omega _{b}^{-}$ & \multicolumn{1}{r}{$%
-0.339$} & \multicolumn{1}{r}{$-0.52$} & --- \\[2pt] 
$\Xi _{bc}^{*0}\leftrightarrow \Xi _{bc}^{0}$ & \multicolumn{1}{r}{$-0.747$}
& \multicolumn{1}{r}{$-1.09$} & --- \\[2pt] 
$\Xi _{bc}^{*0}\leftrightarrow \Xi _{bc}^{\prime \,0}$ & \multicolumn{1}{r}{$%
0.070$} & \multicolumn{1}{r}{$-0.06$} & --- \\[2pt] 
$\Xi _{bc}^{*+}\leftrightarrow \Xi _{bc}^{+}$ & \multicolumn{1}{r}{$0.695$}
& \multicolumn{1}{r}{$1.33$} & --- \\[2pt] 
$\Xi _{bc}^{*+}\leftrightarrow \Xi _{bc}^{\prime \,+}$ & \multicolumn{1}{r}{$%
0.672$} & \multicolumn{1}{r}{$0.95$} & --- \\[2pt] 
$\Omega _{bc}^{*0}\leftrightarrow \Omega _{bc}^{0}$ & \multicolumn{1}{r}{$%
-0.624$} & \multicolumn{1}{r}{$-0.82$} & --- \\[2pt] 
$\Omega _{bc}^{*0}\leftrightarrow \Omega _{bc}^{\prime \,0}$ & 
\multicolumn{1}{r}{$0.112$} & \multicolumn{1}{r}{$0.05$} & --- \\[2pt] 
$\Omega _{bcc}^{*+}\leftrightarrow \Omega _{bcc}^{+}$ & \multicolumn{1}{r}{$%
0.403$} & \multicolumn{1}{r}{$0.42$} & --- \\[2pt] 
$\Xi _{bb}^{*-}\leftrightarrow \Xi _{bb}^{-}$ & \multicolumn{1}{r}{$0.428$}
& \multicolumn{1}{r}{$0.82$} & --- \\[2pt] 
$\Xi _{bb}^{*0}\leftrightarrow \Xi _{bb}^{0}$ & \multicolumn{1}{r}{$-1.039$}
& \multicolumn{1}{r}{$-1.81$} & --- \\[2pt] 
$\Omega _{bb}^{*-}\leftrightarrow \Omega _{bb}^{-}$ & \multicolumn{1}{r}{$%
0.307$} & \multicolumn{1}{r}{$0.52$} & --- \\[2pt] 
$\Omega _{bbc}^{*0}\leftrightarrow \Omega _{bbc}^{0}$ & \multicolumn{1}{r}{$%
-0.395$} & \multicolumn{1}{r}{$-0.42$} & --- \\ \hline\hline
\multicolumn{4}{l}{$^{\ast *}$~Only absolute values $\left| \mu \right| $
are presented.}
\end{tabular}
\end{table}%

We see from Table~\ref{t3.1} that predictions given by the simple
nonrelativistic model (NRQM) and the one with screening and effective quark
mass (SCR) are in almost all cases similar. SCR results are, as a rule,
slightly smaller (not more than $20\%$) than those obtained using NRQM. Only in
two cases (for $\Xi _{c}^{*0}\leftrightarrow \Xi _{c}^{0}$ and $\Xi
_{c}^{*+}\leftrightarrow \Xi _{c}^{\prime \,+}$ transitions) the results
differ significantly, and a large part of this difference comes from the
hyperfine mixing effect. Predictions obtained using the chiral constituent
quark model ($\chi$CQM) are also similar to NRQM results (as a rule, slightly
larger). As expected, predictions for $\Xi _{c}^{*0}\leftrightarrow \Xi
_{c}^{0}$ and $\Xi _{c}^{*+}\leftrightarrow \Xi _{c}^{\prime \,+}$
transitions differ significantly again. Results obtained using the light cone
QCD sum rules (LCSR) for transitions $\Sigma _{c}^{*+}\leftrightarrow
\Lambda _{c}^{+}$, $\Sigma _{c}^{*++}\leftrightarrow \Sigma _{c}^{++}$, $\Xi
_{c}^{*0}\leftrightarrow \Xi _{c}^{0}$, and $\Xi _{c}^{*+}\leftrightarrow
\Xi _{c}^{+}$ are compatible within the error bars with other predictions
($\chi$CQM result for $\Xi _{c}^{*0}\leftrightarrow \Xi _{c}^{0}$ being an
exception). For transitions $\Sigma _{c}^{*0}\leftrightarrow \Sigma
_{c}^{0}$ and $\Sigma _{c}^{*+}\leftrightarrow \Sigma _{c}^{+}$ the LCSR
predictions differ substantially from all other theoretical estimates. Such
difference cannot be understood in the framework of the usual quark model and
therefore looks strange.

Our predictions are, as a rule, smaller than other estimates but on
average closer to results obtained in the nonrelativistic model with
screening and effective quark mass (SCR) -- of course, except
the $\Xi _{c}^{*0}\leftrightarrow \Xi _{c}^{0}$, $\Xi
_{c}^{*+}\leftrightarrow \Xi _{c}^{\prime \,+}$ transitions. In these latter
cases our results agree (at least qualitatively) with NRQM results, because
these transitions are sensitive to the effect of state mixing. The reason
why our results are in general smaller than others is also obvious. In many
cases the contribution of the light quarks to the transition magnetic
moments of heavy baryons is substantial. In the bag model, magnetic moments
of the light quarks residing in heavy hadrons become smaller than in the
light ones \cite{BS13}, and, as a consequence, one obtains relatively
smaller transition moments.

In the bottom sector the transition moments obtained in the framework of the
bag model are smaller than NRQM predictions again, but now they are in good
agreement with LCSR results, while NRQM values are not. This is an
indication that bag model predictions could be treated as serious
improvement over NRQM results. 

Magnetic moments of heavy
quarks are not so sensitive to the environment they live in, and we expect
the transition moments of baryons built up exclusively of heavy quarks to be
similar in all models. Such are triply heavy baryons $\Omega _{bcc}^{+}$ and 
$\Omega _{bbc}^{0}$. Indeed we see from Table~\ref{t3.2} that the bag
model predictions for these baryons differ from NRQM results approximately
by only $5\%$.

Some aspects of our treatment still need some clarification. Maybe the
most interesting question is the role of the colour-hyperfine mixing. The
impact of the hyperfine state mixing on some electromagnetic properties of
heavy baryons has been pointed out in \cite{FLNC81}. Increasing interest in the heavy baryon
spectroscopy made this problem more acute. The extensive study of the
effect of colour-hyperfine mixing on the masses of heavy baryons \cite
{RP08,BS08}, semileptonic decays \cite{RP09,AHN12}, magnetic moments of heavy
baryons \cite{BS13}, and electromagnetic decay rates \cite{AHN10,BFG10} was
performed. The analysis is somewhat complicated by the dependence of 
wave functions on the arrangement of quarks in the spin coupling scheme $%
[[q_{1}q_{2}]^{S}q_{3}]^{J}$ \cite{FLNC81,BS08}. There are three possible
quark ordering schemes. The first is the scheme in which the quarks are
ordered from lightest to heaviest, and the spins of the first two are
coupled to the intermediate spin $S$. Let us call it a light diquark basis.
The second scheme, in which the spins of the lightest and the heaviest
quarks are coupled to the intermediate spin $S$, can be called a heavy-light
diquark basis. The third, in which the two heaviest quarks are coupled to
the intermediate spin, we will call a heavy diquark basis. Strictly speaking,
these notations have little to do with real quark-diquark approximation.
Just convenient names. In order to analyse the dependence on the choice of
quark ordering we have performed calculations in all three ordering schemes.
Expansion coefficients of the physical states in terms of initial wave
functions with definite intermediate spins obtained in these calculations
are presented in Table~\ref{t3.3}.

%t3.3
\begin{table}[tbp] \centering%
\caption{Expansion coefficients of the physical states in terms of wave
functions with definite intermediate spins ($S=0,1$) in three different quark ordering schemes. $q$ stands for the
light quarks ($u$ or $d$).\label{t3.3}} 
\begin{tabular}{lccc}
\hline\hline
Particles & quark ordering & $C_{1}$ & $C_{2}$ \\ \hline
\multicolumn{1}{c}{$\Xi _{c},\Xi _{c}^{\prime \,}$} & \multicolumn{1}{c}{$%
(qs)c$} & \multicolumn{1}{r}{$0.997$} & \multicolumn{1}{r}{$0.073$} \\[4pt] 
\multicolumn{1}{c}{\textquotedbl} & \multicolumn{1}{c}{$(cq)s$} & 
\multicolumn{1}{r}{$-0.562$} & \multicolumn{1}{r}{$0.827$} \\[4pt] 
\multicolumn{1}{c}{\textquotedbl} & \multicolumn{1}{c}{$(sc)q$} & 
\multicolumn{1}{r}{$-0.435$} & \multicolumn{1}{r}{$-0.900$} \\[4pt] 
\multicolumn{1}{c}{$\Xi _{b},\Xi _{b}^{\prime \,}$} & \multicolumn{1}{c}{$%
(qs)b$} & \multicolumn{1}{r}{$0.999$} & \multicolumn{1}{r}{$0.018$} \\[4pt] 
\multicolumn{1}{c}{\textquotedbl} & \multicolumn{1}{c}{$(bq)s$} & 
\multicolumn{1}{r}{$-0.516$} & \multicolumn{1}{r}{$0.857$} \\[4pt] 
\multicolumn{1}{c}{\textquotedbl} & \multicolumn{1}{c}{$(sb)q$} & 
\multicolumn{1}{r}{$-0.484$} & \multicolumn{1}{r}{$-0.875$} \\[4pt] 
\multicolumn{1}{c}{$\Xi _{bc},\Xi _{bc}^{\prime \,}$} & \multicolumn{1}{c}{$%
(qc)b$} & \multicolumn{1}{r}{$0.992$} & \multicolumn{1}{r}{$0.128$} \\[4pt] 
\multicolumn{1}{c}{\textquotedbl} & \multicolumn{1}{c}{$(bq)c$} & 
\multicolumn{1}{r}{$-0.607$} & \multicolumn{1}{r}{$0.795$} \\[4pt] 
\multicolumn{1}{c}{\textquotedbl} & \multicolumn{1}{c}{$(cb)q$} & 
\multicolumn{1}{r}{$-0.385$} & \multicolumn{1}{r}{$-0.923$} \\[4pt] 
\multicolumn{1}{c}{$\Omega _{bc},\Omega _{bc}^{\prime \,}$} & 
\multicolumn{1}{c}{$(sc)b$} & \multicolumn{1}{r}{$0.994$} & 
\multicolumn{1}{r}{$0.112$} \\[4pt] 
\multicolumn{1}{c}{\textquotedbl} & \multicolumn{1}{c}{$(bs)c$} & 
\multicolumn{1}{r}{$-0.593$} & \multicolumn{1}{r}{$0.805$} \\[4pt] 
\multicolumn{1}{c}{\textquotedbl} & \multicolumn{1}{c}{$(cb)s$} & 
\multicolumn{1}{r}{$-0.400$} & \multicolumn{1}{r}{$-0.916$} \\[4pt] \hline\hline
\end{tabular}
\end{table}%

From the table we see that physical states $B$ and $B^{\prime }$
are predominantly light diquark states corresponding to intermediate spin $%
S=0$ and $S=1$, with a small admixture of other ($S=1$ and $S=0$) state. As
expected \cite{FLNC81}, the largest state mixing is seen in the heavy-light
diquark case. The heavy diquark basis is somewhere between light diquark and
heavy-light diquark. In the light diquark basis the mixing is very small.
Nevertheless, as suggested in \cite{FLNC81}, and we have seen in \cite
{BS13}, even such a small mixing can affect the magnetic moments as well as $%
B^{\prime }-B$ transition moments appreciably. So we can anticipate the similar
effect also in the case of $\mathrm{spin\,}\frac{3}{2}\leftrightarrow \frac{1%
}{2}$ transition moments. In Tables~\ref{t3.4} and \ref{t3.5} we compare the
predictions for these transition moments between physical states with
unmixed moments calculated using wave functions corresponding to various
quark ordering schemes. For the singly heavy baryons the results obtained
using heavy-light diquark and heavy diquark schemes are of little interest
(in this case all authors prefer to use the light diquark basis) and are omitted
from the Table~\ref{t3.4}.

%t3.4
\begin{table*}[htb]
\caption{Spin $\frac{3}{2}-\frac{1}{2}$ transition
moments $\mu (B^{*}\leftrightarrow B)$ 
(in nuclear magnetons) of singly heavy baryons for the
physical states ($\Xi _{Q},\Xi _{Q}^{\prime \,}$) and the states with
definite intermediate spins in the light diquark basis.\label{t3.4}} 
\begin{center}
\begin{tabular}{lccccccc}
\hline\hline
State $B$ & $\mu $ & State $B$ & $\mu $ & State $B$ & $\mu $ & State $B$ & $%
\mu $ \\ \hline
\multicolumn{1}{c}{$\Xi _{c}^{0}$} & \multicolumn{1}{r}{$-0.224$} & $\Xi
_{c}^{+}$ & \multicolumn{1}{r}{$1.497$} & $\Xi _{b}^{-}$ & 
\multicolumn{1}{r}{$-0.139$} & $\Xi _{b}^{0}$ & \multicolumn{1}{r}{$1.321$}
\\[2pt] 
\multicolumn{1}{c}{$\Xi _{c}^{\prime \,0}$} & \multicolumn{1}{r}{$-0.915$} & 
$\Xi _{c}^{\prime \,+}$ & \multicolumn{1}{r}{$-0.089$} & $\Xi _{b}^{\prime
\,-}$ & \multicolumn{1}{r}{$-0.415$} & $\Xi _{b}^{\prime \,0}$ & 
\multicolumn{1}{r}{$0.392$} \\[2pt] 
$\left| \lbrack ds]^{0}c\right\rangle $ & \multicolumn{1}{r}{$-0.158$} & 
\multicolumn{1}{r}{$\left| [us]^{0}c\right\rangle $} & \multicolumn{1}{r}{$%
1.507$} & \multicolumn{1}{r}{$\left| [ds]^{0}b\right\rangle $} & 
\multicolumn{1}{r}{$-0.132$} & \multicolumn{1}{r}{$\left|
[us]^{0}b\right\rangle $} & \multicolumn{1}{r}{$1.313$} \\[2pt] 
$\left| \lbrack ds]^{1}c\right\rangle $ & \multicolumn{1}{r}{$-0.931$} & 
\multicolumn{1}{r}{$\left| [us]^{1}c\right\rangle $} & \multicolumn{1}{r}{$%
0.022$} & \multicolumn{1}{r}{$\left| [ds]^{1}b\right\rangle $} & 
\multicolumn{1}{r}{$-0.418$} & \multicolumn{1}{r}{$\left|
[us]^{1}b\right\rangle $} & \multicolumn{1}{r}{$0.416$} \\ \hline\hline
\end{tabular}
\end{center}
\end{table*}

%t3.5
\begin{table}[tbp] \centering%
\caption{Spin $\frac{3}{2}-\frac{1}{2}$ transition
moments $\mu (B^{*}\leftrightarrow B)$ (in nuclear magnetons) of
doubly heavy baryons for the physical (mixed) states and the states in the light diquark, heavy-light
diquark, and heavy diquark schemes.\label{t3.5}} 
\begin{tabular}{lccccc}
\hline\hline
State $B$ & $\mu $ & State $B$ & $\mu $ & State $B$ & $\mu $ \\ \hline
\multicolumn{1}{c}{$\Xi _{bc}^{0}$} & \multicolumn{1}{r}{$-0.747$} & 
\multicolumn{1}{c}{$\Xi _{bc}^{+}$} & \multicolumn{1}{r}{$0.695$} & 
\multicolumn{1}{c}{$\Omega _{bc}^{0}$} & \multicolumn{1}{r}{$-0.624$} \\%
[2pt] 
\multicolumn{1}{c}{$\Xi _{bc}^{\prime \,0}$} & \multicolumn{1}{r}{$0.070$} & 
\multicolumn{1}{c}{$\Xi _{bc}^{\prime \,+}$} & \multicolumn{1}{r}{$0.672$} & 
\multicolumn{1}{c}{$\Omega _{bc}^{\prime \,0}$} & \multicolumn{1}{r}{$0.112$}
\\[2pt] 
$\left| \lbrack dc]^{0}b\right\rangle $ & \multicolumn{1}{r}{$-0.750$} & 
\multicolumn{1}{r}{$\left| [uc]^{0}b\right\rangle $} & \multicolumn{1}{r}{$%
0.604$} & \multicolumn{1}{r}{$\left| [sc]^{0}b\right\rangle $} & 
\multicolumn{1}{r}{$-0.632$} \\[2pt] 
$\left| \lbrack dc]^{1}b\right\rangle $ & \multicolumn{1}{r}{$-0.026$} & 
\multicolumn{1}{r}{$\left| [uc]^{1}b\right\rangle $} & \multicolumn{1}{r}{$%
0.755$} & \multicolumn{1}{r}{$\left| [sc]^{1}b\right\rangle $} & 
\multicolumn{1}{r}{$0.042$} \\[2pt] 
$\left| \lbrack bd]^{0}c\right\rangle $ & \multicolumn{1}{r}{$0.397$} & 
\multicolumn{1}{r}{$\left| [bu]^{0}c\right\rangle $} & \multicolumn{1}{r}{$%
-0.956$} & \multicolumn{1}{r}{$\left| [bs]^{0}c\right\rangle $} & 
\multicolumn{1}{r}{$0.280$} \\[2pt] 
$\left| \lbrack bd]^{1}c\right\rangle $ & \multicolumn{1}{r}{$-0.636$} & 
\multicolumn{1}{r}{$\left| [bu]^{1}c\right\rangle $} & \multicolumn{1}{r}{$%
0.145$} & \multicolumn{1}{r}{$\left| [bs]^{1}c\right\rangle $} & 
\multicolumn{1}{r}{$-0.570$} \\[2pt] 
$\left| \lbrack cb]^{0}d\right\rangle $ & \multicolumn{1}{r}{$0.352$} & 
\multicolumn{1}{r}{$\left| [cb]^{0}u\right\rangle $} & \multicolumn{1}{r}{$%
0.352$} & \multicolumn{1}{r}{$\left| [cb]^{0}s\right\rangle $} & 
\multicolumn{1}{r}{$0.352$} \\[2pt] 
$\left| \lbrack cb]^{1}d\right\rangle $ & \multicolumn{1}{r}{$0.662$} & 
\multicolumn{1}{r}{$\left| [cb]^{1}u\right\rangle $} & \multicolumn{1}{r}{$%
-0.900$} & \multicolumn{1}{r}{$\left| [cb]^{1}s\right\rangle $} & 
\multicolumn{1}{r}{$0.527$} \\ \hline\hline
\end{tabular}
\end{table}%

It is evident that the dependence of unmixed transition moments on the
quark ordering is very strong (see Table~\ref{t3.5}). For example, in the
heavy diquark basis the doubly heavy baryon state $\Xi _{bc}^{\prime \,}$
usually is assumed to be the one with $S=0$. The unmixed transition moments
in this basis then are: $\mu (\Xi _{bc}^{*0}\leftrightarrow \Xi
_{bc}^{\prime \,0}) = \mu (\Xi _{bc}^{*+}\leftrightarrow \Xi _{bc}^{\prime
\,+}) = \mu (\Omega _{bc}^{*0}\leftrightarrow \Omega _{bc}^{\prime
\,0}) = 0.352\,\mu _{N}$. On the other hand, predictions for the physical
states are: $\mu (\Xi _{bc}^{*0}\leftrightarrow \Xi _{bc}^{\prime
\,0}) = 0.070\,\mu _{N}$, $\mu (\Xi _{bc}^{*+}\leftrightarrow \Xi _{bc}^{\prime
\,+}) = 0.672\,\mu _{N}$, and $\mu (\Omega _{bc}^{*0}\leftrightarrow \Omega
_{bc}^{\prime \,0}) = 0.112\,\mu _{N}$.

We see that the best unmixed predictions are obtained in the light diquark
basis. But even this best basis cannot be treated as sufficiently good.
Only for the $\Xi _{b}$, $\Xi _{b}^{\prime }$ states the results are of
rather high accuracy, in almost all other cases the account of the state
mixing effect is important.

From Tables~\ref{t3.1} and \ref{t3.2} we see that some moments are much
smaller than others. Can we find the reason? This is the last point we want
to discuss in this section. By the way, it is a fine example how various
symmetry based considerations work.

Firstly, let us take a look at the $\Xi _{c}^{*0}\leftrightarrow \Xi
_{c}^{0} $ and $\Xi _{b}^{*-}\leftrightarrow \Xi _{b}^{-}$ transitions
forbidden by the \textit{U}-\textrm{spin} symmetry. The \textit{U}-\textrm{%
spin} is similar to isospin in that it is a symmetry in the exchange of $d$
and $s$ quarks, rather than $u$ and $d$ ones. This symmetry connects quarks (%
$d$ and $s)$ with the same charge and therefore is useful for the analysis
of electromagnetic structure of hadrons. Since $d$ and $s$ quarks are the
members of \textit{U}-\textrm{spin }doublet in the case of exact \textit{U}-%
\textrm{spin }symmetry one would have the relation for the reduced magnetic
moments $\bar{\mu}_{d}=\bar{\mu}_{s}$ and analogous relation for magnetic
moments ($\mu _{d}=\mu _{s}$). In turn, the isospin symmetry leads to a
similar relation for the reduced magnetic moments of $u$ and $d$ quarks ($%
\bar{\mu}_{u}=\bar{\mu}_{d}$) but a different relation for quark magnetic
moments (i.\,e., $\mu _{u}=-2\mu _{d}$). The explicit expression for the
transition moment $\mu (\Xi _{c}^{*0}\leftrightarrow \Xi _{c}^{0})$ in terms
of reduced quark magnetic moments is
%3.00
\begin{equation}
\mu (\Xi _{c}^{*0}\leftrightarrow \Xi _{c}^{0})=\frac{\sqrt{2}}{3\sqrt{3}}(%
\bar{\mu}_{s}-\bar{\mu}_{d})\,,  
\label{eq3.00}
\end{equation}
and exactly the same holds for the transition $\Xi _{b}^{*-}\leftrightarrow \Xi
_{b}^{-}$. The \textit{U}-\textrm{spin }conservation would lead to $\mu (\Xi
_{c}^{*0}\leftrightarrow \Xi _{c}^{0})\rightarrow 0$ and $\mu (\Xi
_{b}^{*-}\leftrightarrow \Xi _{b}^{-})\rightarrow 0$. In real world the \textit{U%
}-\textrm{spin} symmetry is broken ($\bar{\mu}_{d}>\bar{\mu}_{s}$), and
these transition moments are to some extent suppressed but not strictly
equal to zero. Moreover, they are enhanced by the hyperfine mixing effect
(especially $\mu (\Xi _{c}^{*0}\leftrightarrow \Xi _{c}^{0})$). Nevertheless,
they still remain smaller than many others.

In the charm sector (see Table~\ref{t3.1}) there are two really very small
transition moments $\mu (\Sigma _{c}^{*+}\leftrightarrow \Sigma _{c}^{+})$
and $\mu (\Xi _{c}^{*+}\leftrightarrow \Xi _{c}^{\prime \,+})$. The
expression for the former assuming isospin symmetry (i.\,e., $\bar{\mu}_{u}=%
\bar{\mu}_{d}$) can be put in the form 
%3.01
\begin{equation}
\mu (\Sigma _{c}^{*+}\leftrightarrow \Sigma _{c}^{+})=\frac{\sqrt{2}}{9}(%
\bar{\mu}_{u}-4\bar{\mu}_{c})\,.  
\label{eq3.01}
\end{equation}

The small value of this transition moment means that an approximate relation 
$\bar{\mu}_{u}\approx 4\bar{\mu}_{c}$ holds. In a framework of the bag model
it looks somewhat accidental. On the other hand, in the naive quark model
the mass of the charmed quark is roughly four times larger than the effective
mass of light quarks. In the nonrelativistic case $\mu _{q}\sim 1/m_{q}$,
and therefore one can expect the value of $(\bar{\mu}_{u}-4\bar{\mu}_{c})$
to be rather small. It is also extra suppressed by the factor $\sqrt{2}/9$.
Note that the usual magnetic moment of the doubly heavy baryon $\Xi
_{cc}^{++}$ given by the expression $\mu (\Xi _{cc}^{++})=\frac{2}{9}(4\bar{%
\mu}_{c}-\bar{\mu}_{u})$ (see Ref.~\cite{BS13}) is also much
smaller than others.

The expression for $\mu (\Xi _{c}^{*+}\leftrightarrow \Xi _{c}^{\prime
\,+})$ is 
%3.02
\begin{equation}
\mu (\Xi _{c}^{*+}\leftrightarrow \Xi _{c}^{\prime \,+})=\frac{\sqrt{2}}{9}(2%
\bar{\mu}_{u}-\bar{\mu}_{s}-4\bar{\mu}_{c})\,.  
\label{eq3.02}
\end{equation}

In the limit of \textit{U}-\textrm{spin} symmetry (the isospin symmetry is
assumed also) $\bar{\mu}_{u}=\bar{\mu}_{d}=\bar{\mu}_{s}$, and Eq.~(\ref
{eq3.02}) becomes equivalent to Eq.~(\ref{eq3.01}). Because the actual $\bar{%
\mu}_{s}$ is smaller than $\bar{\mu}_{u}$, we can expect $\mu (\Xi
_{c}^{*+}\leftrightarrow \Xi _{c}^{\prime \,+})$ to be larger than $\mu
(\Sigma _{c}^{*+}\leftrightarrow \Sigma _{c}^{+})$. This is true for unmixed
moments, however the shift of $\mu (\Xi _{c}^{*+}\leftrightarrow \Xi
_{c}^{\prime \,+})$ due to the hyperfine mixing is negative, and this effect
leads to an opposite relation $\mu (\Xi _{c}^{*+}\leftrightarrow \Xi
_{c}^{\prime \,+})<\mu (\Sigma _{c}^{*+}\leftrightarrow \Sigma _{c}^{+})$.
Since both these transition moments are negative, the absolute value of $\mu
(\Xi _{c}^{*+}\leftrightarrow \Xi _{c}^{\prime \,+})$ is larger.

In the bottom sector there also are two relatively small transition moments, i.\,e., 
$\mu (\Xi _{bc}^{*0}\leftrightarrow \Xi _{bc}^{\prime \,0})$ and $\mu
(\Omega _{bc}^{*0}\leftrightarrow \Omega _{bc}^{\prime \,0})$. The
expressions for these moments in the light diquark basis are
%3.03
\begin{subequations}
\begin{align}
\mu (\Xi _{bc}^{*0}\leftrightarrow \Xi _{bc}^{\prime \,0}) & = \frac{\sqrt{2}}{9}%
(2\bar{\mu}_{c}+2\bar{\mu}_{b}-\bar{\mu}_{u})\, , 
\label{eq3.03a} \\ 
\mu (\Omega _{bc}^{*0}\leftrightarrow \Omega _{bc}^{\prime \,0}) & = \frac{\sqrt{%
2}}{9}(2\bar{\mu}_{c}+2\bar{\mu}_{b}-\bar{\mu}_{s})\,.
\label{eq3.03b}
\end{align}
\end{subequations}

It is evident that transition moments $\mu (\Xi _{bc}^{*0}\leftrightarrow
\Xi _{bc}^{\prime \,0})$ and $\mu (\Sigma _{c}^{*+}\leftrightarrow \Sigma
_{c}^{+})$ have the same order of magnitude. We know that $\bar{\mu}_{b}<%
\bar{\mu}_{c}$. But also $\bar{\mu}_{u}(\Xi _{bc})<\bar{\mu}_{u}(\Sigma
_{c}) $, therefore the difference $\left| \mu (\Xi _{bc}^{*0}\leftrightarrow
\Xi _{bc}^{\prime \,0})\right| -\left| \mu (\Sigma _{c}^{*+}\leftrightarrow
\Sigma _{c}^{+})\right| $ cannot be large. $\mu (\Xi
_{bc}^{*0}\leftrightarrow \Xi _{bc}^{\prime \,0})$ and $\mu (\Omega
_{bc}^{*0}\leftrightarrow \Omega _{bc}^{\prime \,0})$ are also expected to
be of the same order of magnitude because their difference $\frac{\sqrt{2}}{%
9}(\bar{\mu}_{u}-\bar{\mu}_{s})$ vanishes in the \textit{U}-\textrm{spin}
symmetry limit. So we may anticipate the $\mu (\Xi _{bc}^{*0}\leftrightarrow \Xi
_{bc}^{\prime \,0})$ and $\mu (\Omega _{bc}^{*0}\leftrightarrow \Omega
_{bc}^{\prime \,0})$ to be small enough as their partners in the charm sector
were. Both these transition moments undergo positive shifts due to the
hyperfine mixing effect (see Table~\ref{t3.5}), and $\mu (\Xi
_{bc}^{*0}\leftrightarrow \Xi _{bc}^{\prime \,0})$ even changes its sign.
But they still remain smaller than other transition moments.

We have just seen how the light diquark basis facilitates the analysis of
the electromagnetic properties of heavy baryons. Note that in the heavy diquark (as well as heavy-light diquark) basis the suppression of abovementioned transition moments is entirely a hyperfine
mixing effect.

\section{\label{sec_rad}Radiative decay widths}

We wish to end our investigation with the predictions for radiative decay
widths of ground state heavy baryons. We ignore E2 amplitudes which are
expected to be much smaller than M1 transition moments (in the approximation
we are using they are absent). The M1 partial width of the decay $%
B^{*}\rightarrow \gamma \,B$ has the form (see \cite{WBF00}) 
%4.01
\begin{equation}
\Gamma =\frac{\alpha \omega ^{3}}{M_{P}^{2}}\frac{2}{2J+1}\left( \frac{M_{B}%
}{M_{B^{*}}}\right) \mu ^{2}(B^{*}\leftrightarrow B)\,.  
\label{eq4.01}
\end{equation}
Here $\mu (B^{*}\leftrightarrow B)$ is the transition magnetic moment (in
nuclear magnetons), $\alpha =\frac{1}{137}$, $M_{P}$ is the proton mass. $J$
and $M_{B^{*}}$ are the spin and the mass of decaying baryon, $M$ is the
mass of the baryon in its final state, and 
%4.02
\begin{equation}
\omega =(M_{B^{*}}^{2}-M_{B}^{2})/(2M_{B})  
\label{eq4.02}
\end{equation}
is the photon momentum in the c.m. system of decaying baryon.

%t4.1
\begin{table}[tbp] \centering%
\caption{Photon momenta (in MeV) calculated in the
framework of the bag model (Our) and in a nonrelativistic potential model (PM)
compared with an average value $\overline{\omega }=(\omega _{\text{Our}}+\omega _{\text{PM}})/2$ and
with experimental data (Expt.). \label{t4.1}} 
\begin{tabular}{lcccc}
\hline\hline
Decay & Our & PM \cite{RP08} & $\overline{\omega }$ & Expt. \\ \hline
$\Sigma _{c}^{*}\rightarrow \Sigma _{c}$ & \multicolumn{1}{r}{$88$} & 
\multicolumn{1}{r}{$63$} & \multicolumn{1}{r}{$76$} & \multicolumn{1}{r}{$63$%
} \\[2pt]
$\Sigma _{c}^{*}\rightarrow \Lambda _{c}$ & \multicolumn{1}{r}{$184$} & 
\multicolumn{1}{r}{$265$} & \multicolumn{1}{r}{$224$} & \multicolumn{1}{r}{$%
221$} \\[2pt]
$\Sigma _{c}\rightarrow \Lambda _{c}$ & \multicolumn{1}{r}{$99$} & 
\multicolumn{1}{r}{$180$} & \multicolumn{1}{r}{$139$} & \multicolumn{1}{r}{$%
162$} \\[2pt]
$\Xi _{c}^{*}\rightarrow \Xi _{c}$ & \multicolumn{1}{r}{$151$} & 
\multicolumn{1}{r}{$177$} & \multicolumn{1}{r}{$164$} & \multicolumn{1}{r}{$%
169$} \\[2pt]
$\Xi _{c}^{*}\rightarrow \Xi _{c}^{\prime }$ & \multicolumn{1}{r}{$82$} & 
\multicolumn{1}{r}{$54$} & \multicolumn{1}{r}{$68$} & \multicolumn{1}{r}{$67$%
} \\[2pt]
$\Xi _{c}^{\prime }\rightarrow \Xi _{c}$ & \multicolumn{1}{r}{$72$} & 
\multicolumn{1}{r}{$125$} & \multicolumn{1}{r}{$98$} & \multicolumn{1}{r}{$%
105$} \\[2pt]
$\Omega _{c}^{*}\rightarrow \Omega _{c}$ & \multicolumn{1}{r}{$75$} & 
\multicolumn{1}{r}{$57$} & \multicolumn{1}{r}{$66$} & \multicolumn{1}{r}{$72$%
} \\[2pt]
$\Sigma _{b}^{*}\rightarrow \Sigma _{b}$ & \multicolumn{1}{r}{$30$} & 
\multicolumn{1}{r}{$25$} & \multicolumn{1}{r}{$27$} & \multicolumn{1}{r}{$21$%
} \\[2pt]
$\Sigma _{b}^{*}\rightarrow \Lambda _{b}$ & \multicolumn{1}{r}{$155$} & 
\multicolumn{1}{r}{$241$} & \multicolumn{1}{r}{$198$} & \multicolumn{1}{r}{$%
209$} \\[2pt]
$\Sigma _{b}\rightarrow \Lambda _{b}$ & \multicolumn{1}{r}{$123$} & 
\multicolumn{1}{r}{$216$} & \multicolumn{1}{r}{$170$} & \multicolumn{1}{r}{$%
188$} \\[2pt]
$\Xi _{b}^{*}\rightarrow \Xi _{b}$ & \multicolumn{1}{r}{$124$} & 
\multicolumn{1}{r}{$171$} & \multicolumn{1}{r}{$148$} & \multicolumn{1}{r}{$%
152$} \\ \hline\hline
\end{tabular}
\end{table}%

In our calculations we have used transition moments from the preceding
section for the $\mathrm{spin\,}\frac{3}{2}\leftrightarrow \frac{1}{2}$
decays and transition moments obtained in our earlier paper \cite{BS13} for 
$B^{\prime }\rightarrow B$ decays. At present, in the absence of
experimental data, we see no reliable way to estimate possible uncertainties
of calculated transition moments and use them as they are. Another source of
errors in the calculation of decay widths is the uncertainty in the value
of photon momentum (\ref{eq4.02}). The problem we are encountered with is
that bag model predictions for baryon masses and corresponding mass
differences are not of very high quality. Nevertheless, some regularities
exist. One can
check that the bag model almost always overestimates the baryon mass
difference of $B^{*}-B^{\prime }$ type. For example, such are $\Sigma
_{c}^{*}-\Sigma _{c}$, $\Xi _{c}^{*}-\Xi _{c}^{\prime }$, $\Omega
_{c}^{*}-\Omega _{c}$, $\Sigma _{b}^{*}-\Sigma _{b}$ (see Table~\ref{t4.1},
values of photon momentum presented in this table do not differ significantly
from the corresponding mass differences). Furthermore, the remaining baryon
mass differences of $B^{*}-B$ and $B^{\prime }-B$ type are, as a rule,
underestimated. An opposite tendency is seen in results
obtained using the nonrelativistic potential model \cite{RP08} (see Table~\ref
{t4.1} again). To
our knowledge this is the only paper that includes a full list of theoretical
predictions for the masses of ground state heavy baryons we need. In order to minimize the uncertainties in the
calculation of photon momenta we will use the experimental masses of baryons
if available. The corresponding
momenta are presented in Table~\ref{t4.1} (column Expt.) For all remaining transitions we use a somewhat arbitrary
prescription which serves rather well in many cases where
experimental masses are known. In the cases when
experimental data are absent, our proposal is to use an average of our bag model and
before-mentioned potential model result, $\overline{\omega }=(\omega
_{\text{Our}}+\omega _{\text{PM}})/2$. To justify this choice, we compare in Table~\ref
{t4.1} the bag model predictions for $\omega $ with potential model \cite
{RP08} predictions, average momenta $\overline{\omega }$, and experimental
momenta (Expt.) calculated using experimental values of the baryon masses.
The mass of $\Xi _{b}^{*\,}$ is taken from \cite{CMS12}, all others from
Particle Data Tables \cite{PDG12}.

%t4.2
\begin{table}[tbp] \centering%
\caption{Radiative decay widths (in keV) of charmed baryons.\label{t4.2}} 
\begin{tabular}{ccccc}
\hline\hline
Decay & Our & Bag \cite{HDDT78} & RQM \cite{BFG10,IKLR99} & LCSR \cite{AAO09}
\\ \hline
$\Sigma _{c}^{*\,0}\rightarrow \Sigma _{c}^{0}$ & $1.08$ & $2.67$ & --- & $%
0.08\pm 0.03$ \\[4pt] 
$\Sigma _{c}^{*\,+}\rightarrow \Sigma _{c}^{+}$ & $0.004$ & $1.52$ & $%
0.14\pm 0.004$ & $0.40\pm 0.16$ \\[4pt] 
$\Sigma _{c}^{*\,+}\rightarrow \Lambda _{c}^{+}$ & $126$ & $176.7$ & $151\pm
4$ & $130\pm 45$ \\[4pt] 
$\Sigma _{c}^{+}\rightarrow \Lambda _{c}^{+}$ & $46.1$ & $22.91$ & $60.7\pm
1.5$ & --- \\[4pt] 
$\Sigma _{c}^{*\,++}\rightarrow \Sigma _{c}^{++}$ & $0.826$ & $3.27$ & --- & 
$2.65\pm 1.20$ \\[4pt] 
$\Xi _{c}^{*\,0}\rightarrow \Xi _{c}^{0}$ & $0.908$ & --- & $0.68\pm 0.04$ & 
$0.66\pm 0.32$ \\[4pt] 
$\Xi _{c}^{*\,0}\rightarrow \Xi _{c}^{\prime \,0}$ & $1.03$ & --- & --- & ---
\\[4pt] 
$\Xi _{c}^{\prime \,0}\rightarrow \Xi _{c}^{0}$ & $0.0015$ & --- & $0.17\pm
0.02$ & --- \\[4pt] 
$\Xi _{c}^{*\,+}\rightarrow \Xi _{c}^{+}$ & $44.3$ & $74.01$ & $54\pm 3$ & $%
52\pm 25$ \\[4pt] 
$\Xi _{c}^{*\,+}\rightarrow \Xi _{c}^{\prime \,+}$ & $0.011$ & $1.46$ & ---
& --- \\[4pt] 
$\Xi _{c}^{\prime \,+}\rightarrow \Xi _{c}^{+}$ & $10.2$ & --- & $12.7\pm
1.5 $ & --- \\[4pt] 
$\Omega _{c}^{*\,0}\rightarrow \Omega _{c}^{0}$ & $1.07$ & $0.85$ & --- & ---
\\[4pt] 
$\Xi _{cc}^{*\,+}\rightarrow \Xi _{cc}^{+}$ & $2.08$ & $3.96$ & $28.79\pm
2.51$ & --- \\[4pt] 
$\Xi _{cc}^{*\,++}\rightarrow \Xi _{cc}^{++}$ & $1.43$ & $4.35$ & $23.46\pm
3.33$ & --- \\[4pt] 
$\Omega _{cc}^{*\,+}\rightarrow \Omega _{cc}^{+}$ & $0.949$ & $1.35$ & $%
2.11\pm 0.11$ & --- \\[4pt] \hline\hline
\end{tabular}
\end{table}%

%t4.3
\begin{table*}[htb]
\caption{Radiative decay widths (in keV) of singly heavy bottom
 baryons.\label{t4.3}} 
\begin{center}
\begin{tabular}{l@{~~~}c@{~~~}c@{~~~}cc}
\hline\hline
Transition & Our & LCSR \cite{AAO09} & LCSR-2$^{**}$ & HQET \cite{ZD99}$%
^{**} $ \\ \hline
$\Sigma _{b}^{*-}\rightarrow \Sigma _{b}^{-}$ & $0.010$ & $0.11\pm 0.06$ & $%
0.0076\pm 0.0041$ & $0.020$ \\[2pt] 
$\Sigma _{b}^{*0}\rightarrow \Sigma _{b}^{0}$ & $0.005$ & $0.028\pm 0.016$ & 
$0.0017\pm 0.0009$ & $0.0051$ \\[2pt] 
$\Sigma _{b}^{*0}\rightarrow \Lambda _{b}^{0}$ & $81.1$ & $114\pm 45$ & $%
84.5\pm 33.4$ & $254$ \\[2pt] 
$\Sigma _{b}^{0}\rightarrow \Lambda _{b}^{0}$ & $58.9$ & --- & --- & $194$ \\%
[2pt] 
$\Sigma _{b}^{*+}\rightarrow \Sigma _{b}^{+}$ & $0.054$ & $0.46\pm 0.22$ & $%
0.030\pm 0.014$ & $0.080$ \\[2pt] 
$\Xi _{b}^{*-}\rightarrow \Xi _{b}^{-}$ & $0.278$ & $1.50\pm 0.75$ & $%
0.464\pm 0.232$ & --- \\[2pt] 
$\Xi _{b}^{*-}\rightarrow \Xi _{b}^{\prime \,-}$ & $0.005$ & --- & --- & ---
\\[2pt] 
$\Xi _{b}^{\prime \,-}\rightarrow \Xi _{b}^{-}$ & $0.118$ & --- & --- & ---
\\[2pt] 
$\Xi _{b}^{*0}\rightarrow \Xi _{b}^{0}$ & $24.7$ & $135\pm 65$ & $41.4\pm
20.6$ & --- \\[2pt] 
$\Xi _{b}^{*0}\rightarrow \Xi _{b}^{\prime \,0}$ & $0.004$ & --- & --- & ---
\\[2pt] 
$\Xi _{b}^{\prime \,0}\rightarrow \Xi _{b}^{0}$ & $14.7$ & --- & --- & --- \\%
[2pt] 
$\Omega _{b}^{*-}\rightarrow \Omega _{b}^{-}$ & $0.006$ & --- & --- & --- \\ 
\hline\hline
\multicolumn{5}{l}{$^{\ast *}$~Results obtained using current data for the
masses of heavy baryons.}
\end{tabular}
\end{center}
\end{table*}

%t4.4
\begin{table}[tbp] \centering%
\caption{Radiative decay widths (in keV) of doubly and triply heavy bottom
baryons.\label{t4.4}} 
\begin{tabular}{lccc}
\hline\hline
Transition & Our & RQM \cite{BFG10} & PM \cite{AHN10} \\ \hline
$\Xi _{bc}^{*0}\rightarrow \Xi _{bc}^{0}$ & $0.612$ & $0.51\pm 0.06$ & $1.03$
\\[2pt] 
$\Xi _{bc}^{*0}\rightarrow \Xi _{bc}^{\prime \,0}$ & $0.0003$ & $(2\pm
2)\times 10^{-6}$ & $0.0012$ \\[2pt] 
$\Xi _{bc}^{\prime \,0}\rightarrow \Xi _{bc}^{0}$ & $0.125$ & $0.31\pm 0.04$
& $0.209$ \\[2pt] 
$\Xi _{bc}^{*+}\rightarrow \Xi _{bc}^{+}$ & $0.533$ & $0.46\pm 0.10$ & $%
0.739 $ \\[2pt] 
$\Xi _{bc}^{*+}\rightarrow \Xi _{bc}^{\prime \,+}$ & $0.031$ & $0.0015\pm
0.0007$ & $0.061$ \\[2pt] 
$\Xi _{bc}^{\prime \,+}\rightarrow \Xi _{bc}^{+}$ & $0.037$ & $0.14\pm 0.03$
& $0.124$ \\[2pt] 
$\Omega _{bc}^{*0}\rightarrow \Omega _{bc}^{0}$ & $0.239$ & $0.29\pm 0.03$ & 
$0.502$ \\[2pt] 
$\Omega _{bc}^{*0}\rightarrow \Omega _{bc}^{\prime \,0}$ & $0.0005$ & $(1\pm
1)\times 10^{-6}$ & $0.0031$ \\[2pt] 
$\Omega _{bc}^{\prime \,0}\rightarrow \Omega _{bc}^{0}$ & $0.053$ & $0.21\pm
0.02$ & $0.085$ \\[2pt] 
$\Omega _{bcc}^{*+}\rightarrow \Omega _{bcc}^{+}$ & $0.004$ & --- & --- \\%
[2pt] 
$\Xi _{bb}^{*-}\rightarrow \Xi _{bb}^{-}$ & $0.022$ & $0.059\pm 0.014$ & ---
\\[2pt] 
$\Xi _{bb}^{*0}\rightarrow \Xi _{bb}^{0}$ & $0.126$ & $0.31\pm 0.06$ & --- \\%
[2pt] 
$\Omega _{bb}^{*-}\rightarrow \Omega _{bb}^{-}$ & $0.011$ & $0.0226\pm
0.0045 $ & --- \\[2pt] 
$\Omega _{bbc}^{*0}\rightarrow \Omega _{bbc}^{0}$ & $0.005$ & --- & --- \\ 
\hline\hline
\end{tabular}
\end{table}%

We see that the averaged momentum in most cases is a significant improvement
over the bag model predictions ($\Omega _{c}^{*}\rightarrow \Omega _{c}$ decay
being an exception) and also over the potential model results (with the
exception for the $\Sigma _{c}^{*}\rightarrow \Sigma _{c}$ and $\Sigma
_{b}^{*}\rightarrow \Sigma _{b}$ decays). Therefore we expect that for other
baryons (when data are absent) the averaged momentum also ought to be a
reasonable choice. To calculate the kinematical factor $M_{B}/M_{B^{*}}$ we
again use experimental masses when available. If experimental data are
absent, we resort to the bag model results. As a consequence, our results for
the radiative decay widths are not pure bag model predictions. But we think
these improved results should be more accurate and therefore more useful.
They are presented in Tables~\ref{t4.2}--\ref{t4.4}. We also compare our
predictions with results obtained using several other approaches. These are:

\begin{itemize}
\item  Earlier MIT bag model predictions \cite{HDDT78} (Bag).

\item  Nonrelativistic potential model \cite{AHN10} (PM).

\item  Relativistic three-quark model \cite{BFG10,IKLR99} (RQM).

\item  Light cone QCD sum rules \cite{AAO09} (LCSR).

\item  LCSR estimates in the leading order of heavy quark effective theory 
\cite{ZD99} (HQET) calculated using current data for the masses of heavy
baryons.
\end{itemize}

For charmed baryons (Table~\ref{t4.2}) the results obtained in all approaches
form a varying pattern. In general most of them are more or less compatible.
Nevertheless, there are exceptions. One can single out the LCSR prediction
for $\Sigma _{c}^{*\,0}\rightarrow \Sigma _{c}^{0}$ decay width, which is
an order of magnitude smaller than others, and a fussy mess-up in the $\Sigma
_{c}^{*\,+}\rightarrow \Sigma _{c}^{+}$ decay rates predicted using various
approaches. Another outstanding difference is predictions for the $\Xi
_{cc}^{*\,+}\rightarrow \Xi _{cc}^{+}$ and $\Xi _{cc}^{*\,+}\rightarrow \Xi
_{cc}^{+}$ decay widths obtained in the relativistic three-quark model \cite
{BFG10}. These are an order of magnitude larger than our predictions.

On the other hand, we predict very small decay widths for the M1 transitions 
$\Sigma _{c}^{*\,+}\rightarrow \Sigma _{c}^{+}$, $\Xi _{c}^{\prime
\,0}\rightarrow \Xi _{c}^{0}$, and $\Xi _{c}^{*\,+}\rightarrow \Xi
_{c}^{\prime \,+}$. By the way, $\Sigma _{c}^{*\,+}\rightarrow \Sigma _{c}^{+}$ and 
$\Xi _{c}^{*\,+}\rightarrow \Xi _{c}^{\prime \,+}$ are the very same decays the small transition moments of which were discussed in detail in
the preceding section.

In the case of singly heavy bottom baryons (Table~\ref{t4.3}), when comparing our
results with LCSR \cite{AAO09} predictions we are faced with an astonishing
disagreement. Since transition moments in both approaches agree well (see
Table~\ref{t3.2}), we guess that in Ref.~\cite{AAO09} the obsolete data
for experimental baryon masses have been used. Therefore we recalculated these
decay widths using the same transition moments (given in Ref.~\cite{AAO09})
but with updated experimental values of baryon masses (in the calculation of
photon momentum). The results are presented in the column denoted as LCSR-2
of Table~\ref{t4.3}. As expected, the corrected LCSR predictions are in
satisfactory agreement with our results.

For doubly heavy $B_{bc}$ and $B_{bb}$ baryons (see Table~\ref{t4.4}) our
predictions are compatible (at least qualitatively) with the estimates of
radiative decay rates obtained using nonrelativistic potential model (PM) 
\cite{AHN10} and relativistic three-quark model (RQM) \cite{BFG10}. For
example, all three models predict small decay widths for the transitions $%
\Xi _{bc}^{*0}\rightarrow \Xi _{bc}^{\prime \,0}$ and $\Omega
_{bc}^{*0}\rightarrow \Omega _{bc}^{\prime \,0}$. Note that in all these
approaches the state mixing due to colour-hyperfine interaction was taken into
account. Our predictions for the decays $\Xi _{bc}^{*0}\rightarrow \Xi
_{bc}^{0}$, $\Xi _{bc}^{*0}\rightarrow \Xi _{bc}^{\prime \,0}$, $\Xi
_{bc}^{*+}\rightarrow \Xi _{bc}^{+}$, $\Xi _{bc}^{*+}\rightarrow \Xi
_{bc}^{\prime \,+}$, and $\Omega _{bc}^{*0}\rightarrow \Omega _{bc}^{\prime
\,0}$ are somewhere between decay rates obtained in PM and RQM. For other
transitions we predict somewhat smaller decay widths. In Table~\ref{t4.4}
we also give estimates for the decay rates of triply heavy baryons $\Omega
_{bcc}^{*+}\rightarrow \Omega _{bcc}^{+}$ and $\Omega _{bbc}^{*0}\rightarrow
\Omega _{bbc}^{0}$. They are relatively small, because small are the
corresponding photon momenta $\omega $.

\section{\label{sec_sum}Summary}

Using the modified bag model employed before in the study of magnetic moments of
heavy baryons \cite{BS13} we have
analysed the radiative decays of baryons containing one, two, and three heavy
quarks. All heavy baryons are treated
on the same footing. We have calculated the M1 transition moments for all
ground state heavy baryons. These transition moments were used to obtain
predictions for partial decay rates. To our knowledge for some transitions ($%
\Xi _{b}^{*-}\rightarrow \Xi _{b}^{\prime \,-}$, $\Xi _{b}^{\prime
\,-}\rightarrow \Xi _{b}^{-}$, $\Xi _{b}^{*0}\rightarrow \Xi _{b}^{\prime
\,0}$, $\Xi _{b}^{\prime \,0}\rightarrow \Xi _{b}^{0}$, $\Omega
_{b}^{*-}\rightarrow \Omega _{b}^{-}$, $\Omega _{bcc}^{*+}\rightarrow \Omega
_{bcc}^{+}$, and $\Omega _{bbc}^{*0}\rightarrow \Omega _{bbc}^{0}$) it is
the first theoretical estimate. 
In the case of baryons containing three quarks of different flavours the
state mixing due to colour-hyperfine interaction was taken into account.
Because so far there are no experimental data to compare our predictions
with, we have compared our results for magnetic transition moments and decay
rates with those obtained in various other theoretical approaches. In many
cases a good agreement was found. The existing differences were pointed
out and in some cases the possible source of discrepancy was discussed.

\end{document}